\documentclass[]{article}

\usepackage{arxiv}

\usepackage[utf8]{inputenc} % allow utf-8 input
\usepackage[T1]{fontenc}    % use 8-bit T1 fonts
\usepackage{hyperref}       % hyperlinks
\usepackage{url}            % simple URL typesetting
\usepackage{booktabs}       % professional-quality tables
\usepackage{amsfonts}       % blackboard math symbols
\usepackage{nicefrac}       % compact symbols for 1/2, etc.
\usepackage{microtype}      % microtypography
\usepackage{lipsum}
\usepackage{graphicx}
\usepackage{natbib}
\usepackage{index, amsmath, listings, multirow, nomencl, ltablex, longtable,  rotating, scalerel,  setspace, subfig,tabularx, titlesec,url, verbatim, adjustbox, algpseudocode, algorithmicx, algorithm}

\graphicspath{Figures/}

\title{Comparison of Parametric versus Machine-learning Multiple Imputation in Clinical Trials with Missing Continuous Outcomes}

\author{
Mia S. Tackney \\
  MRC-Biostatistics Unit, University of Cambridge, Cambridge, United Kingdom  \\
  \texttt{mst35@cam.ac.uk} \\
   \And
 Jonathan W. Bartlett  \\
  Department of Medical Statistics, London School of Hygiene and Tropical Medicine, London, UK 
  \And
 Elizabeth Williamson \\
  Department of Medical Statistics, London School of Hygiene and Tropical Medicine, London, UK
  \And 
  Kim May Lee \\
Institute of Psychiatry, King's College London, London, UK}

\begin{document}
\maketitle
\begin{abstract}
The use of flexible machine-learning (ML) models to generate imputations of missing data within the framework of Multiple Imputation (MI) has recently gained traction, particularly in non-randomised observational settings. For randomised controlled trials (RCTs), it is unclear whether ML approaches to MI provide valid inference, and whether they outperform parametric MI approaches under some complex data generating mechanisms. We conducted two simulation studies in RCT settings that have incomplete continuous outcomes but fully observed covariates. We compared Complete Cases, standard MI (MI-norm), MI with predictive mean matching (MI-PMM) and ML-based approaches to MI, including classification and regression trees (MI-CART), Random Forests (MI-RF) and SuperLearner when outcomes are missing completely at random or missing at random conditional on treatment/covariate. The first simulation explored a cross-sectional outcome with non-linear covariate-outcome relationships in the presence/absence of covariate-treatment interactions. The second simulation explored skewed repeated measures, motivated by a trial with digital outcomes. In the absence of interactions, we found that Complete Cases yields reliable inference; MI-norm performs similarly, except when missingness depends on the covariate. ML approaches can lead to smaller mean squared error than Complete Cases and MI-norm in specific non-linear settings, but provide unreliable inference for others. Imputation by MI-PMM  with the default implementation in the \texttt{R} package \texttt{mice} can lead to unreliable inference under some settings. In the presence of complex treatment-covariate interactions, performing MI separately by arm, either with MI-norm, MI-RF or MI-CART, provides inference that has comparable or with better properties compared to Complete Cases when the analysis model omits the interaction term. For ML approaches, we observed that unreliable inference arises in terms of bias in the estimated effect and/or its standard error when Rubin’s Rules are implemented.   
\end{abstract}

% keywords can be removed

\section{Introduction}
Missing data is common in Randomised Clinical Trials (RCTs) and can arise due to a number of reasons, such as missed clinical visits, participant drop-out, and practical issues such as difficulty in obtaining laboratory test results. While monitoring and corrective actions, such as reminder calls or escalation of missing lab results, can reduce the proportion of missing data, it is difficult to prevent missing data entirely \citep{Jakobsen2017}. If handled incorrectly in the analysis, missing data can lead to biased and imprecise estimates of treatment effects \citep{NRC2010}. \\

Multiple Imputation (MI) is a popular approach to handling missing data \citep{Kenward2007, Schafer1999} with several widely used software implementations \citep{Hayati2015, Mackinnon2010}. We focus on the RCT setting where the outcome is continuous and has missing values, while the treatment assignment and covariates are fully observed. In this setting, Complete Case analysis of a normal linear regression model for an outcome at a single time point is unbiased if the data are Missing at Random (MAR) given the model's covariates and the model is correctly specified \citep{Sullivan2018}. However, MI has several potential advantages in the setting where missingness occurs in the outcome only. Firstly, it allows important auxiliary variables to be incorporated in the imputation model, which can make the MAR assumption more plausible, without necessitating changes to the analysis model. Secondly, MI allows sensitivity analysis to be performed within a coherent framework, which is critically important for assessing the impact of violations of the untestable MAR assumption \citep{Cro2016}. \\

Traditionally, MI has been based on specification of a parametric imputation model for the incomplete variable(s). In non-randomised settings, there has been increased interest in using Machine Learning (ML) approaches, such as classification and regression trees (CART), k-nearest neighbours (kNN) or SuperLearners, to perform imputation within the MI framework \citep{Shah2014, Burgette2010, Stekhoven2012}. A particular advantage of these methods is their ability to accommodate complex relationships between variables in the imputation without the need for the user to explicitly specify the form of relationships or parametric distributions. These methods are clearly attractive in the analysis of large scale epidemiological studies, where analysis models can be complex with a high number of variables with potential non-linearities and interactions. In RCTs, where pre-specification of analyses are required, including pre-specification of missing data methods, the potential for ML approaches in MI to guarantee good performance in the presence of unknown non-linearities or interactions, could be a major advantage. However, potential advantages and drawbacks of ML approaches to MI have not been explored in the RCT setting. In particular, whether ML approaches to MI provide valid inferences in terms of Type I error control and confidence interval coverage has not been explored; this is particularly important for pivotal drug trials, as regulatory authorities require strict Type I error control \citep{ICH1998E9}. \\

The aim of this study is to compare default implementations of MI with ML approaches to MI in clinical trial settings where continuous outcomes are missing. In most statistical software, the default implementation of MI is based on a parametric model with linear additive effects of covariates and treatment (on some scale). We consider trial settings where the analysis model is a linear regression of the outcome on treatment and baseline covariate, but these parametric assumptions may be violated and there is potential for ML approaches to provide benefit. Specifically, we consider settings where there are: 

\begin{itemize}
    \item non-linear relationships between the covariate and cross-sectional outcome and absence/presence of a complex treatment-covariate interaction;
    \item skewness in repeatedly measured outcomes.
\end{itemize}

The article is structured as follows. In Section \ref{section_ML}, we describe the general framework of MI and introduce ML approaches to MI which have been proposed in the literature. In Section \ref{Section_compare}, we provide an overview of studies which compared ML and default approaches to MI in non-randomised settings. In Section \ref{Section_sim1and2}, we provide a simulation study comparing standard and ML approaches to MI in trial settings where there are (i) non-linearities in the covariate-outcome relationship and (ii) interactions between treatment and outcome. Section \ref{Section_sim_3} provides a simulation study based on the 2019 iNO-PF trial which has repeated measures and a skewed outcome. Finally, Section \ref{Section_discussion} provides discussion of the results and points to important areas of future work. \\

\section{Multiple Imputation}
\label{section_ML}
A range of imputation approaches have been proposed to handle missing data. Single imputation, including mean imputation and regression imputation, refers to the process of generating a single imputed dataset. This does not reflect uncertainty due to the missing data and treats imputed values as if they were observed in the analysis, leading to underestimated standard errors (SEs) if the imputed dataset is analysed using standard methods without allowance for the imputation step \citep{LittleRubin2019, Sterneb2393}. Conversely, Multiple Imputation (MI), as described below, accounts for uncertainty, and under certain conditions, provides valid variance estimates.  \\

Consider an RCT setting where $n$ participants are individually randomised to a binary treatment. We describe MI for our setting where the treatment assignment and covariates are fully observed but the outcome can have missing values. We denote by $Y$ the $n$-vector for the continuous outcome, and denote by $X$ the matrix containing $n$-vectors for $p$ continuous covariates: $X= \{X_1, ..., X_p\}$. The $n$-vector $Z$ is for treatment assignment, with entries in $\left\{0, 1 \right\}$ indicating assignment to control and experimental treatment, respectively.  We denote the outcome and covariates for participants assigned to treatment $z \in \left\{0,1 \right\}$ as $Y^z, X_1^z, ..., X_p^z$. Further, we denote by $Y^{z, obs}$ and $Y^{z, mis}$ the observed and missing components of $Y^z$.  \\

 We assume that the analysis model is a linear regression of the outcome on treatment and a pre-specified subset of the covariates, $X'=\{X_1, ..., X_r\}$, with $r\leq p$. This is a common choice of analysis models in the RCT setting. The remaining covariates $\{X_{r+1}, ..., X_p\}$ are included in the imputation process as auxiliary variables to make the MAR assumption more plausible. \\

We describe MI conducted separately by trial arm, as this is often recommended for RCTs \citep{White2011, Sullivan2018}. This is closely related to performing MI on the full dataset with a single imputation model which includes interaction terms between the treatment and all other variables. The latter explicitly assumes that the residual error variances are equal across arms, whereas imputing separately by arm relaxes this assumption.

\begin{algorithm}[H]
\caption{Multiple Imputation by treatment arm}
\label{MI_alg}
\begin{itemize}
    \item[] Step 1: \textit{Impute}\\
    Separately by treatment arm $Z \in \left\{0, 1 \right\}$, use a normal linear regression to model the observed component of the outcome $Y^{Z, obs}$ on the covariates $X_1^{Z}, ..., X_p^{Z}$. Unless otherwise specified, we assume that the covariates are included additively and linearly in the regression model. The missing values $Y^{Z, mis}$ are replaced by simulated draws from the posterior predictive distribution of $Y^{Z}$. Non-informative priors are assumed for the coefficients of covariates (assumed to be continuous) and the residual variance. \\

     We repeat the \textit{Impute} step $M$ times to create $M$ imputations. \\

    \item[] Step 2: \textit{Analyse}\\
    For each $m \in \left\{1, 2, ..., M \right\}$, we fit the following linear model to the data from both arms, which consists of imputed and observed outcomes: 

    \begin{equation}
    \label{analysis_model}
        \mathbb{E}(Y \mid Z, X^r) = \beta_0 + \beta_1 Z + \beta_2 X_1 + ... + \beta_{r+1} X_r.
    \end{equation}

     The treatment effect $\beta_1$ is typically the parameter of interest in clinical trials.\\
     
     \item[] Step 3 \textit{Pool}: \\
     We pool results from the $M$ regressions via Rubin's rules \citep{Rubin1987} to get a pooled estimate of $\beta_1$ and its standard error, which reflects uncertainty due to the missing data.  
    \end{itemize}
\end{algorithm}

In Step 1 of Algorithm \ref{MI_alg}, imputations are drawn from the posterior predictive distribution. We refer to this approach as \textit{MI-norm}. An alternative approach is predictive mean matching \citep{Little1988}, which we refer to as \textit{MI-PMM}.  In PMM, to impute missing values for $Y^Z$, predicted values are generated for $Y^Z$ with the linear imputation model specified in Step 1. For each missing observation in $Y^Z$, a donor pool is created by selecting observed values whose predicted values are closest to that of the missing observation. One donor is randomly drawn from the donor pool to replace the missing value. PMM has become the default option in some statistical packages, such as the \texttt{R} package \texttt{mice} (see Table \ref{R_Code}). Unless otherwise stated,  \textit{MI-norm} and \textit{MI-PMM} include covariates additively and linearly in the imputation model; the analysis model is assumed to also include treatment and covariates additively and linearly.

\subsection{ML approaches to Multiple Imputation}

We now describe how the Multiple Imputation procedure can be modified to incorporate Machine-learning approaches to imputation. Instead of specifying a parametric model in Step 1 to generate imputations, ML approaches learn about relationships between variables via data-driven algorithms. These algorithms usually require specification of \textit{tuning parameters} which control model complexity; these parameters affect the model performance and are typically specified by the user or chosen via cross-validation \citep[p. 222]{Hastie2009}. For each ML approach, we describe the algorithm in general (e.g. for prediction), then describe how it is used for single imputation and MI, and highlight key tuning parameters required. 

\subsubsection{k-Nearest Neighbours (kNN) }
K-nearest neighbors (kNN) is an ML algorithm where predictions are generated by taking a weighted mean of $k$ observed values identified as ``nearest neighbours" according to a distance metric calculated using the covariates in the imputation model. The most commonly used distance metric for continuous variables is the Minkowski distance, which includes Euclidean and Manhattan distances as special cases. The number of neighbours, $k$, is the tuning parameter. \\

 kNN was first used for single imputation of continuous variables by \cite{Troyanskaya2001}. Several adaptations to single imputation via kNN have been proposed to make the method more widely applicable. \cite{Zhang2012} proposed distance metrics which handle combinations of continuous and categorical variables, both as predictors and as variables to be imputed. Weighted distance metrics were proposed by \citep{Faisal2017}, which take into account correlations between the variable to be imputed and predictors. Further, 
\citep{Kim2004} developed a sequential approach to single imputation with kNN, which imputes variables starting with those that have the lowest proportion of missingness. Imputed values are then used as predictors in the imputation of other variables. \\

Multiple Imputation via kNN imputation was proposed by \cite{Faisal2021} using a weighted distance metric, where weights are determined by the correlation between predictors and the variable to be imputed. MI via kNN begins with an initial cross-validation step where two tuning parameters are learned: one is an exponent which controls how correlations contribute to the distance metric, and the second controls how those correlations are mapped into weights in the distance metric. Variables with missing values are imputed sequentially, in the order of their proportion of missingness. For the first variable, available data is used to compute distance calculations where predictors have missing values. For subsequent variables, previous imputations are used to compute distances.  \\

\subsubsection{Classification and Regression Trees (CART)}
Classification and Regression Trees (CARTs) generate predictions by partitioning the outcome variable, recursively splitting the data into two groups based on values of the predictors. We focus on Regression trees, which are used for continuous outcomes, as opposed to Classification trees, which are used for categorial outcomes. At each iteration, a predictor is chosen and a split point is selected which best separates the outcome. The split point is typically chosen to minimise the outcome variable sum of squares calculated within each split. This process results in a partition of the individuals into relatively homogeneous groups with respect to the outcome variable \citep[p. 305]{Hastie2009}.  Tuning parameters in CARTs include the \textit{minimum leaf size} ({minimum number of observations) required in the final grouping to prevent overfitting, and \textit{complexity parameter} (threshold for improvement in model fit); further splits are stopped once this threshold is reached.  \\

\cite{Burgette2010} proposed an MI approach via CARTs (MI-CART) within the chained equations framework \citep{vanBuuren2011}. In MI-CART, in order to initially fill in missing values, a CART for each variable with missing values is fit to the observed data and this tree is used to generate initial imputations. Then, CARTs are used instead of parametric regression models to generate imputations variable-by-variable. Draws from the predictive distribution are generated by identifying the grouping (leaf) that matches the predictors, given the CART, and randomly selecting a value from the corresponding leaf. In order to attempt to reflect uncertainty about the estimated distribution around the missing outcome (i.e. making the imputation `proper'), \cite{Burgette2010} used a bootstrapping procedure within each leaf before drawing imputations. \\

\subsubsection{Random Forests (RF)}
Motivated by the desire to reduce the risk of overfitting with single classification or regression trees, Random Forests (RF) were developed. These fit multiple CARTs and combine their predictions  \citep{Breiman2001}. A RF proceeds by generating bootstrap resamples from the dataset and training a CART independently in each bootstrap resample. At each split of each the CART, a random subset of predictors are used instead of the full set of predictors. The CARTs are then aggregated by taking the average of predictions (for a continuous outcome) or the mode (for a categorical outcome). The \textit{number of trees} to grow, as well as the \textit{random input selection} (number of predictors considered at each split) are tuning parameters for RFs. \\
 
 \cite{Stekhoven2012} developed Missforest, an RF algorithm for single imputation which can handle a mixture of continuous and categorial variables as predictors. Multiple Imputation via RF (MI-RF) was then developed by \cite{Shah2014} via chained equations. In MI-RF, initial values are generated for missing data via mean or mode imputation. Then, a bootstrap sample is obtained; this helps to reflect uncertainty about the estimated distribution. An RF is trained iteratively for each variable with missing data in the sample; this involves a further bootstrap step where a decision tree is fit to each bootstrap resample with a specified number of predictors to consider at each split. Imputed values are drawn from independent normal distributions with the mean given by the averaged of predicted values across all trees and residual variance given by the ``out of bag" mean squared error (the mean of squared differences between each observed value and prediction generated by trees which omit the observation in the bootstrap resample).\\

\subsubsection{SuperLearner}
A SuperLearner is an ensemble algorithm which generates a predictive model using a weighted combination of user-specified models, which can include any number of semi-parametric or non-parametric models (e.g. linear models, Random Forests, kNN) \citep{Naimi2018}. The weights are selected to minimise a specified loss function for the imputation model via cross-validation, such as mean-squared error. SuperLearners lead to asymptotic prediction errors that are at least as low as those provided by the best-performing algorithm in the ensemble \citep{vanDerLaan2007}. \\

MI via SuperLearners (MI-SuperLearner) has been proposed by  \cite{Laqueur2021} and \cite{Carpenito2022} using the chained equations framework.  MI-SuperLearner begins by generating initial values for missing data via mean or mode imputation. A bootstrap resample of the data is generated and a SuperLearner model is fitted. Imputations are generated from a normal distribution with mean given by the SuperLearner prediction, where out-of-fold predictions from each model in the ensemble are combined using the optimised ensemble weights; residual variance is given by a local kernel estimate \citep{Laqueur2021}. \\

Table \ref{R_Code} summarises the available \texttt{R} packages to implement the MI approaches we have described, together with default settings and tuning parameters. 

\begin{table}[H]
\caption{R packages and tuning parameters for MI approaches}
\centering
\small
\begin{tabular}{|l|l|l|}
\hline
\textbf{MI Approach} 
& \textbf{R Implementation} 
& \textbf{Tuning Parameters and default settings} 
\\ \hline

MI-norm 
& \texttt{mice} with \texttt{method = "norm"} 
& ---

\\ \hline

MI-PMM 
& \texttt{mice} with \texttt{method="PMM"} (default) 
& \begin{tabular}[c]{@{}l@{}}
\texttt{donors} (number of donors) = 5

\end{tabular}
\\ \hline

MI-kNN 
& \begin{tabular}[c]{@{}l@{}}
No open source R implementation\\ for multiple imputation. \\
Single imputation available \\via \texttt{VIM} and \texttt{DMwR} packages.
\end{tabular}
& --- 
\\ \hline

MI-CART 
& \texttt{mice} with \texttt{method = "cart"} 
& \begin{tabular}[c]{@{}l@{}}
\texttt{minbucket} (minimum leaf size) = 5\\
\texttt{cp} (complexity parameter) = $10^{-4}$

\end{tabular}
\\ \hline

MI-RF 
& \texttt{mice} with \texttt{method = "rf"} 
& \begin{tabular}[c]{@{}l@{}}
\texttt{ntree} (number of trees) = 10\\
\texttt{mtry} (random input selection) = $\sqrt{p}$\\
\end{tabular}
\\ \hline

MI-SuperLearner 
& \texttt{SuperMICE} and \texttt{misl} packages 
& Depends on choice of learners 
\\ \hline
\end{tabular}
\label{R_Code}
\end{table}

\section{Comparisons between parametric and Machine-Learning approaches to Multiple Imputation}
\label{Section_compare}

In this section we describe existing literature which compared parametric MI-norm or MI-PMM (where a linear model is assumed by default) to ML approaches to MI in non-randomised settings. To our knowledge, no such comparison has been made in the setting of randomised studies.  \\

\cite{Burgette2010}  compared MI-CART with MI-norm in simulations where the outcome was generated with interaction terms and quadratic terms as predictors. In these simulations, while the imputation model for MI-norm included predictors linearly and additively, the analysis model included the interaction and quadratic terms. For parameters in the regression model for this outcome they identified poor confidence interval coverage using these methods, but overall lower bias and higher precision from using MI-CART. \cite{Shah2014} compared MI-norm, MI-PMM and MI-RF in a simulation study based on epidemiological settings where the analysis of interest was a multivariable Cox. Both the imputation model and analysis model included a quadratic term (for age) but no interactions between predictors. All MI procedures led to estimates of log-hazard ratios with little to no bias, but MI-RF produced narrower confidence intervals. While MI-norm and MI-PMM led to slight undercoverage ($92-94\%$), MI-RF lead to slightly higher standard errors and overcoverage ($97-98\%$). Further, they conducted a simulation with a non-linear relationship between the variable with missing data and a fully observed variable. MI-RF led to reduced bias and coverage between $94\%$ and $97.5\%$, while MI-norm led to increased bias and lower coverage of $79.8\%$. A greater number of trees led to increased precision for continuous outcomes at the cost of computation time and increased bias. For categorical variables, performance was similar across selected values of the number of trees. \\

\cite{Doove2014} extended the simulation study by \cite{Burgette2010} and compared the default implementation of MI (with PMM) to MI-CART, MI-RF with bootstrapping, and MI-RF with bootstrapping and random input selection. In their simulation, a continuous outcome was generated with two-way interactions and quadratic terms as predictors. They considered a range of interaction types, including varying degrees of correlation between variables that interact. The imputation models included predictors additively and linearly while the analysis model included the interactions and quadratic terms. They found that the interaction terms were estimated with less bias and coverage closer to $95\%$ using MI-CART compared to MI-PMM and MI-RF, particularly where the effect size is small. However, for some main effects, MI-PMM appeared to perform best with lowest bias and greater coverage. They also investigated a categorial variable setting and again found that MI-RF led to improved estimates of interaction terms.\\  

\cite{Dashti2024} compared parametric MI-PMM to MI-RF in a setting where the Average Treatment Effect (ACE) is estimated using Targeted Maximum Likelihood Estimation (TMLE). Their simulation based on an epidemiological setting compared MI-PMM approaches with varying levels of interactions in the imputation model (none, one-way, two-way and three-way interactions) with MI-CART and MI-RF. They also considered non-MI approaches such as Complete Cases and extensions of TMLE. They concluded that MI-PMM generally outperformed MI-RF and MI-CART, except in more complex settings where there are quadratic terms. They found that MI-RF consistently had large bias and did not perform as well as MI-CART, as was shown in \cite{Doove2014}.\\

\cite{Laqueur2021} compared  MI-PMM, MI-norm  and MI-SuperLearner in a simulation study where data were generated under a (i) quadratic relationship, (ii) lognormal relationship, (iii) zero-inflated Poisson distribution and (iv) a binary outcome with induced MAR patterns. MI-SuperLearner used mean imputation, imputation via linear model, locally estimated scatterplot smoothing (LOESS) (for continuous variables), generalized additive model, neural networks and linear discriminant analysis (for binary variables). They found that MI-SuperLearner had lower bias and coverage closer to $95\%$ overall compared to MI-norm and MI-PMM, particularly when over $30\%$ of variables had missing values.  Further, \cite{Carpenito2022} compared parametric MI-PMM, MI-CART, MI-RF and MI-SuperLearner in a simulation study in simulation settings explored by \cite{Doove2014}. They found that MI-SuperLearner had bias closest to zero and smallest confidence width compared to other approaches. Coverage by MI-SuperLearner was higher than other approaches, displaying overcoverage for some coefficients and slight undercoverage for other coefficients. \\

 Overall, in non-randomised observational settings, ML approaches to MI have been shown to have improved performance, particularly in terms of bias, over MI-PMM or MI-norm in settings where there are (i) non-linearities, (ii) interactions and (iii) skewness in outcomes. However, confidence interval coverage was not always correct; the extent to which ML approaches to MI have correct frequentist properties needs further investigation. Furthermore, the performance of ML approaches in RCT settings is unknown, where correct confidence interval coverage is critical (particularly for pivotal trials) and analysis models are typically simpler than in observational studies.

\section{Simulation Study: Non-linearities and interactions}
\label{Section_sim1and2}
\subsection{Aim}
The aim of this simulation study is to compare the performance of (i) standard MI (ii) MI with with ML approaches and (iii) Complete case analysis in estimating the treatment effect in clinical trial settings. We focus on scenarios with increasingly extreme non-linearities and interactions in order to be able to identify settings where more flexible imputation methods may provide benefit.\\

We consider two settings. In the first setting, there is a single covariate (plus randomised treatment) and a non-linear relationship between the covariate and outcome. In the second setting, there are treatment-covariate interactions. In both settings, outcomes are missing with MCAR or MAR mechanisms. There are no missing values in the covariate.

\subsection{Data-Generating Mechanisms}

\subsubsection{Single covariate setting without interaction}
We generated a continuous outcome from the model, 
\begin{equation}
\label{data_gen}
Y_i = \beta_0 + \beta_1 Z_i + f(X_i) + \epsilon_i,
\end{equation}
where $i \in \left\{1,...,n\right\}$,  $\epsilon_i \sim N(0, 42)$, and the binary treatment $Z_i$ takes value 1 for the active arm and 0 for the placebo arm. Treatment was randomly assigned with a probability of 0.5. We considered the case with a treatment effect ($\beta_1=40$) and without treatment effect ($\beta_1=0$).  The covariate was generated according to $f(X_i)$, where $X_i$ is drawn from a $N(0,1)$ distribution, and the function $f(\cdot)$ denotes five possible covariate--outcome relationships: \textit{linear, two-tier, flattening, quadratic} and \textit{harmonic}, as illustrated in Figure \ref{true_relationships_rev}. Full details are provided in Supplementary File 1. 

\begin{figure}[H]
\centering
\includegraphics[width=0.8\textwidth]{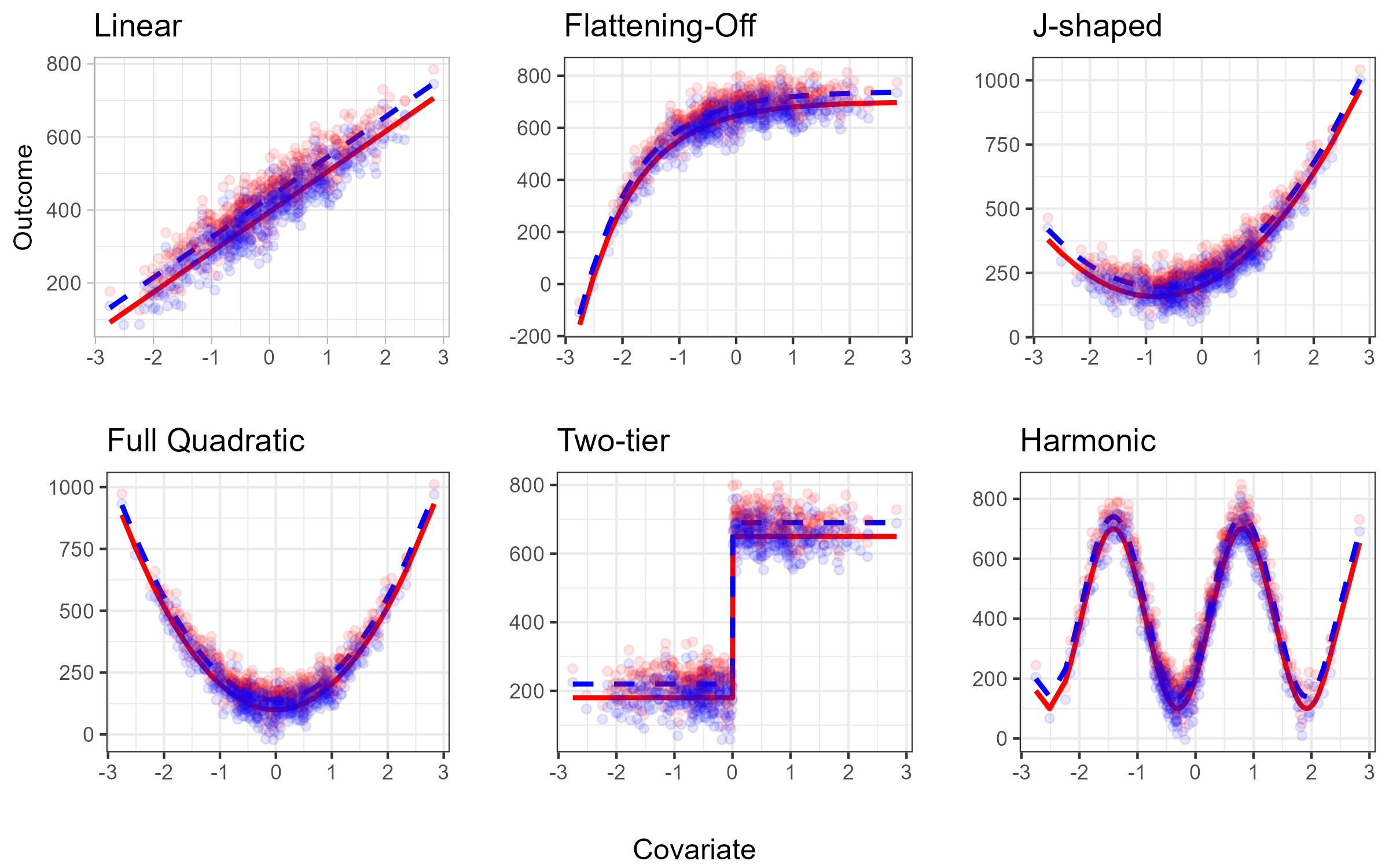}
\caption{True relationships between the continuous outcome and treatment studied in the simulation with the dots depicting the datapoints in a single simulated dataset of 500 participants. Red indicates the active treatment and blue indicates the control.}
\label{true_relationships_rev}
\end{figure}

\subsubsection{Interaction setting}

Four different interaction settings were considered, illustrated in Figure \ref{true_interactions_rev}, each featuring a continuous outcome, one covariate and a covariate--treatment interaction. A single covariate was generated from a $N(0,1)$ distribution. In the first scenario, the covariate has a small interaction with the treatment. In the second scenario, the covariate has a larger interaction in which the treatment effect changes direction. In the third scenario, the covariate--outcome relationship has different shapes in each arm (exponential under the active treatment and linear under the placebo). Finally, in the last scenario, the covariate is the square of a standard normally distributed variable and, therefore, has a skewed distribution. In this last setting, the covariate affects the outcome only for the control arm. Full details are provided in Supplementary File 1.
\begin{figure}[H]
\centering
\includegraphics[width=1\textwidth]{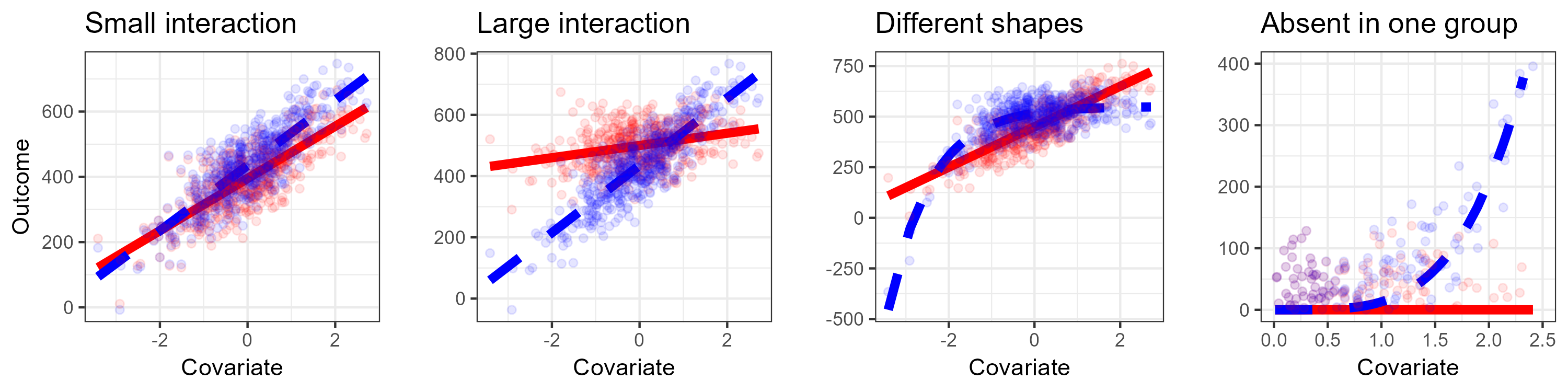}
\caption{True relationships between the continuous covariate and outcome with interaction are shown: small interaction, large
interaction, different shapes, and effect absent in one group. The dots depict the datapoints in a single simulated dataset of 500 participants, where
the red dots indicate the active treatment and the blue dots indicate the control.}
\label{true_interactions_rev}
\end{figure}

In all settings, we consider sample sizes of $n=$50, 100, 200, and 500. 

\subsubsection{Missing data Mechanisms}

The missingness indicator $R_i$ was generated as $R_i \sim \mbox{Bernoulli(}\phi_i)$, where 

\begin{equation}
\label{missing_eq}
\mbox{logit}(\phi_i) = \alpha_0 + \alpha_1 X_i + \alpha_2 Z_i. 
\end{equation}

The parameters $\alpha_0, \alpha_1$ and $\alpha_2$ govern the overall quantity of missing data and the extent to which missingness depends on the covariate and treatment. We consider the following missing data mechanisms: Missing Completely at Random (MCAR), Missing at Random depending on baseline value $X_i$ and missing at random depending on treatment $Z_i$. We consider mechanisms that lead to $10 \%$ and $30\%$ of the sample being missing. Full details of the mechanisms are in Table \ref{miss_mech}.

\begin{table}[H]
\centering
\caption{Details of the Missing Data Mechanisms: total percentage of missing outcomes, parameter values in Equation \eqref{missing_eq}, and brief description.}
\label{miss_mech}
\begin{tabular}{|l|l|l|l|l|l|}
\hline
\textbf{Mechanism} & \textbf{\begin{tabular}[c]{@{}l@{}}Missing \\ outcomes \\ (\%)\end{tabular}} & $\alpha_0$ & $\alpha_1$ & $\alpha_2$ & \textbf{Description}                                                                                             \\ \hline
MCAR               & 10                                                                            & -2.197     & 0          & 0          & $10\%$ of the sample is missing.                                                                                 \\ \hline
MCAR               & 30                                                                            & -0.847     & 0          & 0          & $30\%$ of the sample is missing.                                                                                 \\ \hline
MAR given X        & 10                                                                            & -2.5       & -1         & 0          & \begin{tabular}[c]{@{}l@{}}80\% of individuals with missing\\  outcomes have $x_i < 0$.\end{tabular}            \\ \hline
MAR given X        & 30                                                                            & -1         & -1         & 0          & \begin{tabular}[c]{@{}l@{}}$74\%$ of individuals with missing \\ outcomes have $x_i<0$.\end{tabular}            \\ \hline
MAR given Z        & 10                                                                            & -1.55      & 0          & -1.9       & \begin{tabular}[c]{@{}l@{}}$85\%$ of individuals with missing \\ outcomes receive the control arm.\end{tabular} \\ \hline
MAR given Z        & 30                                                                            & -0.4       & 0          & -1         & \begin{tabular}[c]{@{}l@{}}$67\%$ of individuals with missing \\ outcomes receive the control arm.\end{tabular} \\ \hline
\end{tabular}
\end{table}

\subsection{Estimand}
The estimand of interest is the average treatment effect. We estimate this using a linear regression which adjusts for the treatment and covariate (Equation \eqref{analysis_model} with $p=1$ covariate). 
While this model is mis-specified when there are non-linear covariate-outcome relationships and interactions, it is a common choice of analysis model in trial settings with a continuous outcome and has been shown, under 1:1 randomisation and complete data, to provide consistent point estimates \citep{Yang2001} and standard errors of the treatment effect \citep{Wang2019}.

\subsection{Missing data and analysis methods}
\label{sim_methods}

We considered methods for handling missing data listed below. Across 
all MI methods, we impute separately by treatment arm and use $M=30$ imputations. The analysis model is a normal linear regression model adjusting for a covariate and treatment effect, as in Equation \eqref{analysis_model} with $p=1$ covariate.

\begin{itemize}
    \item \textit{Complete cases (CC)}
    \item \textit{MI with normal linear imputation (MI-norm)} \\
    A linear relation between the covariate and outcome is specified. No interactions are included. Imputations are generated via Bayesian linear regression with conjugate priors for the regression parameters (non-informative normal prior for regression coefficients and inverse-gamma prior for the residual variance). This is implemented using the \texttt{norm} option in \texttt{mice}.
    \item \textit{MI with predictive mean matching (MI-PMM)}\\
   A linear relation between the covariate and outcome is specified.  No interactions are included. Imputations are generated via predictive mean matching with 5 donors. This is implemented using the default \texttt{PMM} option in \texttt{mice}.
    \item \textit{MI with classification and regression trees (MI-CART)} \\
    Imputations are generated via  the \textit{CART} option in \texttt{mice}. The complexity parameter, \texttt{cp}, is set to $10^{-4}$, and we explore the following values of minimum leaf size (\textit{minbucket}): 5 (the default), 10 and 20.
    \item \textit{MI with random forest (MI-RF)} \\
    Imputations are generated via the \texttt{RF} option in \texttt{mice}. We explore the following values of the number of trees (\textit{ntree}): 5, 10 (the default) and 20. 
    \item \textit{MI with SuperLearner (MI-SL)}\\
    Imputations are generated via an ensemble of the following learners: mean model, GAMs, GLMs and randomForest using the \texttt{SuperMICE} package. 
    MI-SL is omitted for $n=500$ for all covariate-outcome relationships due to computational time. Further, MI-SL is omitted for the Harmonic relationship for $n=200$ due to computational time. 
\end{itemize}

\subsection{Performance Measures}
\label{performance_measures}

We repeat the data generating and analysis process 5000 times and summarise the simulation outputs using the following performance measures:
\begin{itemize}
\item Bias of the estimated treatment effect
\item Coverage of the $95\%$ confidence interval
\item Ratio of the mean estimated SE of treatment effect estimate (from the analysis model) to empirical SE of treatment effect estimate.
\item Mean Squared Error (MSE) of the estimated treatment effect 
\item Type I error (for scenarios under the Null)
\item Power (for scenarios under the Alternative)
\end{itemize} 

The ratio of the estimated SE of the treatment effect estimate is particularly useful for assessing whether the variability estimated using Rubin's Rules adequately captures the total variability. A ratio of SE $<1$ indicates that the sampling variance of the treatment effect estimate is underestimated, while a ratio of SE $>1$ indicates overestimation.  As coverage mirrors Type I error, coverage is omitted in plots displaying Null scenarios. \\

There were some instances where the analysis method failed to produce sensible results. Extreme results were occasionally detected with MI-SuperLearner; we excluded replications when computing the performance measures when estimated treatment effect exceeded 200 or was less than -200. Further, for some repetitions with sample size  50, several methods failed to produce treatment effect estimates. Table 1 in Supplementary File 1 reports the number of failed replications for simulation settings where over $0.5\%$ repetitions failed. \\

The computational time required for one repetition of each simulation setting is presented in Table 2 in Supplementary File 1. 

\subsection{Results}

\subsubsection{Single covariate setting} 
Figure \ref{Null_200} displays results for the single covariate setting when $n=200$ and there is no treatment effect. We show results of missing data mechanisms where $30\%$ of data are missing. The tuning parameter, \textit{minbucket} for RF appeared not to affect the performance measures; therefore, we display results with the default value of 5. For CART, we display results when \textit{ntree} set to 5; poorer performance was observed for \textit{ntree} set to 10 and 20 (see GitHub repository). \\

We observe that Complete Cases leads to favourable performance overall: estimates have minimal bias and model-based SEs approximate the empirical SEs well across all relationships. Type I error is also controlled at approximately $5\%$ across all covariate-outcome relationships. These may be explained by the fact that under the MCAR, MAR-X and MAR-Z mechanisms, the joint distribution of the covariate and outcome is identical in the two treatment arms in the Complete Cases (of reasonable size) when there is null effect, such that there is zero average treatment effect in the (expected) Complete Cases subset. Therefore, under the null hypothesis the point estimate and standard error provided by the analysis model in Equation \eqref{analysis_model} are consistent despite misspecification of the covariate-outcome relationship \citep{Wang2019,Bartlett2020}. See Supplementary File 3 for a proof.\\

MI-norm performed similarly to Complete Cases under the MCAR and MAR-Z mechanisms. While there may be considerable mismatch between the true covariate-outcome relationship and the linear relationship specified in the imputation model, the average treatment effect is estimated without bias. However, when missingness depends on baseline covariate, for some non-linear relationships, MI-norm led to under-estimated model-based SEs and hence inflated Type I error rate. \\

MI-PMM led to bias in the estimated treatment effect in several settings, particularly when missingness depended on treatment. It also led to inflated model-based SEs for the Harmonic relationship and under-estimated model-based SEs for the Full Quadratic relationship, resulting in deflated and inflated type I error rates, respectively. These results may be a consequence of the misspecified imputation model; for PMM, the predicted values may be incorrectly matched and the selected donors are further from the true outcome than expected.\\

The ML approaches (MI-RF, MI-CART and MI-SL) perform similarly to each other: for non-linear relationships, they often result in bias when missingness depends on treatment group, but otherwise result in negligible bias. For relationships that are linear or closely approximated by a linear relationship, we observe under-estimated model-based SEs for these approaches, with considertably greater bias for MI-SL. As a result, the Type I error rate is inflated above the $5\%$ nominal level, particularly when missingness depends on baseline covariate (and for most scenarios for MI-SL). Nevertheless, for the full quadratic, two-tier and harmonic relationships, we observe that MSE is lower for the ML approaches compared to Complete Cases, MI-norm and MI-PMM. We see similar patterns for other sample sizes; the reduced MSE for ML methods becomes apparent for sample sizes 100 and above (see Supplementary File 2). \\

\begin{figure}[]
\centering
\includegraphics[width=1\textwidth]{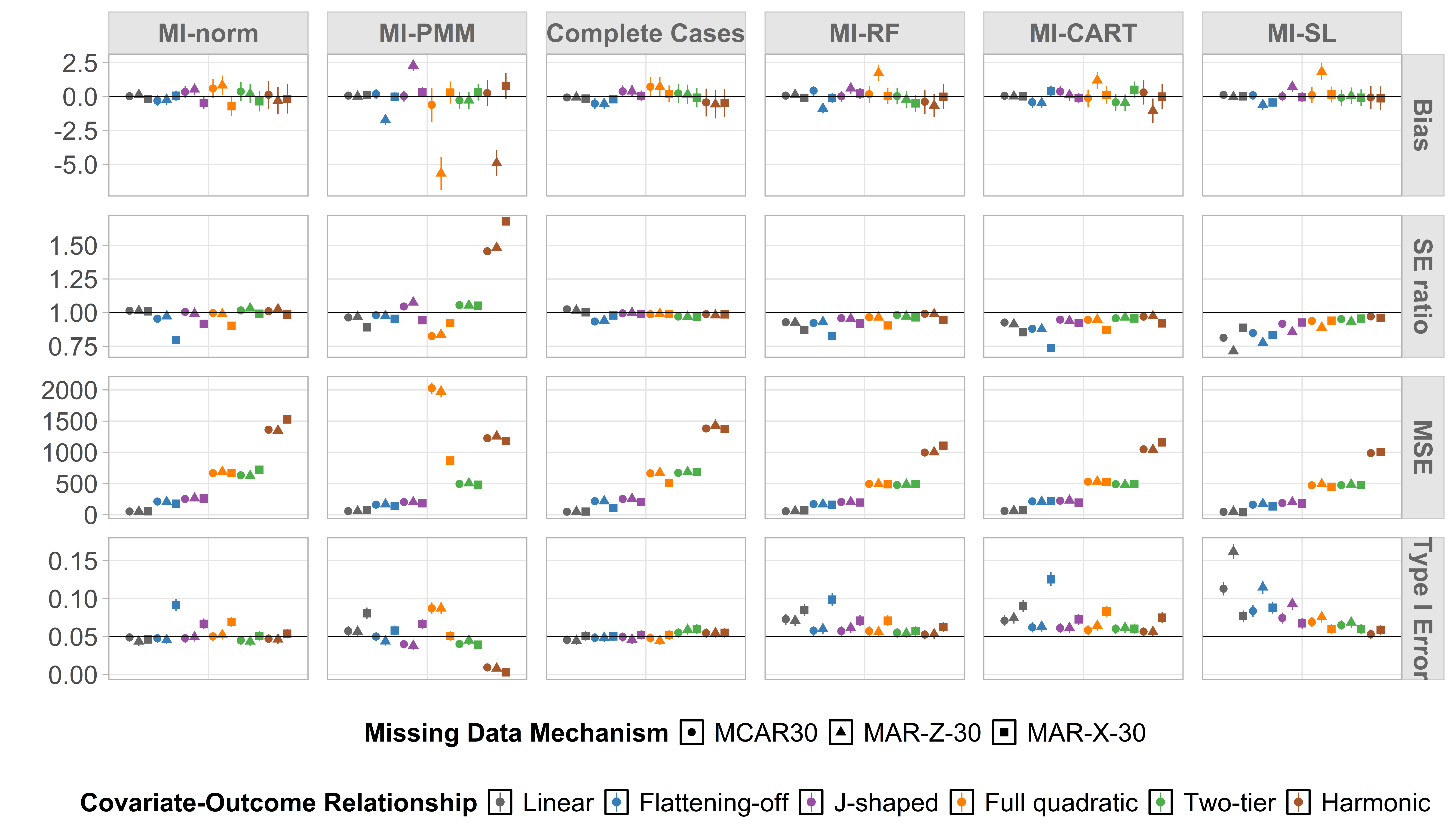}
\caption{ Simulation results for the single covariate, no interaction setting for sample size 200 when the true treatment effect is 0. Performance of missing data methods in terms of bias, ratio of model-based versus empirical standard error and Type I error are shown. Note: for RF, the number of trees is set to 5 and for CART, minbucket is set to 5. The Harmonic relationship with MI-SL was omitted due to computational time.  Estimates are indicated with $\pm 1.96 \times $ Monte Carlo error bars (but are often too small to be seen due to the scale of the plots).  }\label{Null_200}
\end{figure}

We display results when $n=200$ when there is a treatment effect in Figure \ref{Alternative_200}. For Complete Cases, MI-norm and MI-PMM, we draw similar conclusions to those obtained under the null scenario. \\

For ML approaches when sample sizes are 100 and above, we note that MSE is lower for the ML methods compared to MI-norm, MI-PMM and Complete cases for the full quadratic, two-tier and harmonic relationships. Further, for sample sizes 200 and above, MI-RF and MI-CART have an advantage over Complete Cases and MI-norm in the Two-tier relationship and Harmonic relationship under MCAR30 or MAR-Z-30; they lead to a slight increase in power and a slight reduction standard error while maintaining coverage close to $95\%$.  For other relationships, the model SE underestimates the empirical SE when MI-RF and MI-CART are implemented, leading to undercoverage, particularly when missingness depends on baseline covariate. MI-SL has lower coverage than all other methods. \\

\begin{figure}[]
\centering
\includegraphics[width=1\textwidth]{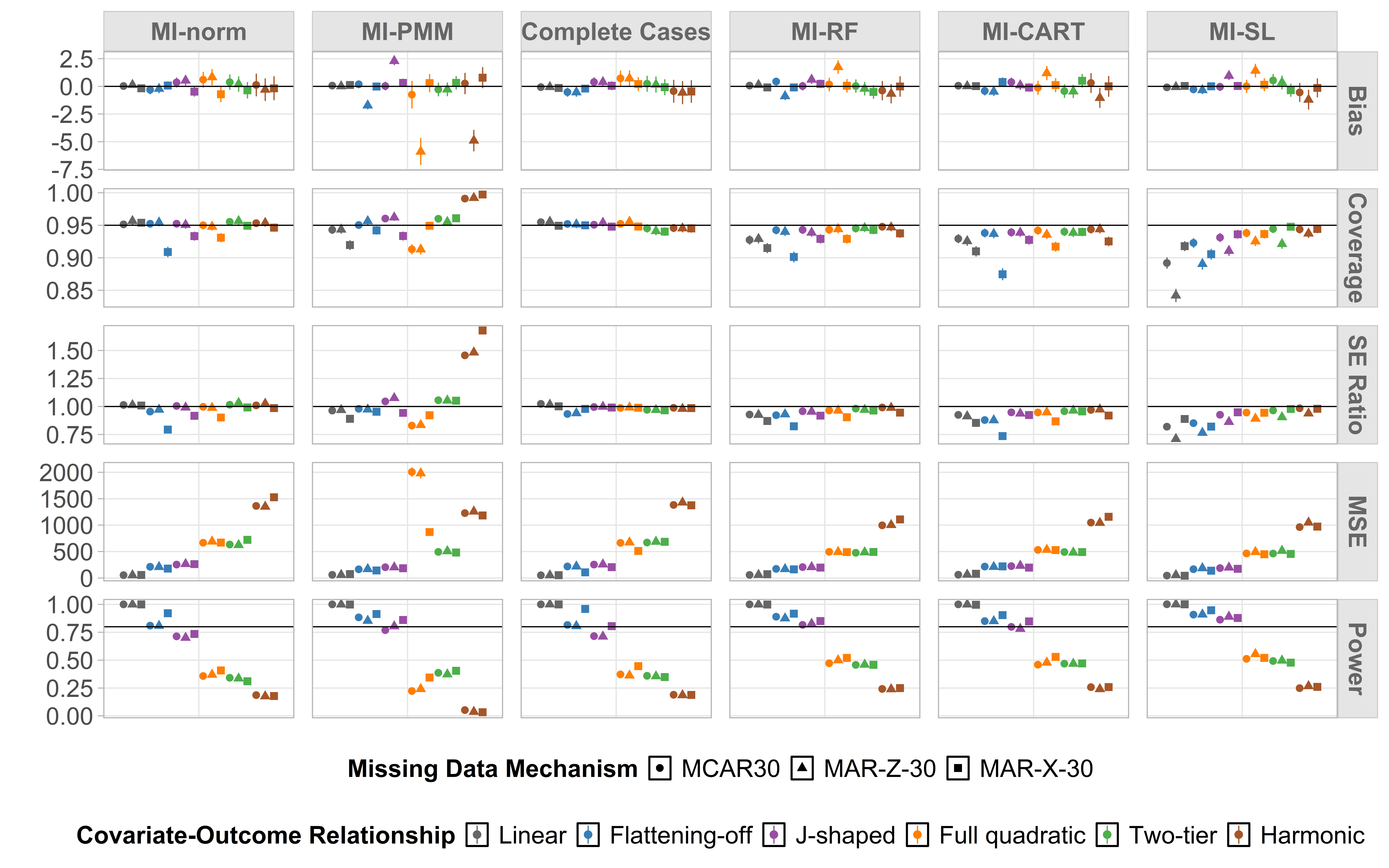}
\caption{Simulation results for the single covariate, no interaction setting for sample size 200 when the true treatment effect is 40. Performance of missing data methods in terms of bias, ratio of model-based versus empirical standard error and power are provided across seven different missing data mechanisms. Note: for RF, the number of trees is set to 5 and for CART, minbucket is set to 5. Estimates are indicated with $\pm 1.96 \times $ Monte Carlo error bars (but are often too small to be seen due to the scale of the plots).   }
\label{Alternative_200}
\end{figure}

\subsubsection{Interaction Setting}
Figure \ref{interaction_200} displays results for the interaction setting under the alternative scenario when $n=200$ and $30\%$ of outcomes are missing. Again, we display results where \textit{minbucket} is set to 5 for RF and \textit{ntree} is set to 5 for CART. \\

For Complete Cases, estimates for the average treatment effect are unbiased under MCAR and MAR-Z. However, we observe bias under MAR-X and undercoverage. In the interaction setting, the true conditional treatment effect varies with $X$. Since the distribution of the covariate in the Complete Cases is different from that of the full randomised sample, we observe bias in the treatment effect; see Supplementary File 3 for a proof.  In the Absent in One Arm setting, we observe undercoverage under MCAR and MAR-Z in addition to MAR-X, and we observe under-estimated model-based SEs under MCAR and MAR-X. \\

For MI-norm and MI-PMM, we observe bias in the Different Shapes and Absent in One Arm setting where missingness depends on baseline covariate. In both settings, the covariate-outcome relationship is different in the two arms; therefore, the imputation model is mis-specified in both arms in distinct ways. Further, we observe bias,  under-estimated model SEs and hence undercoverage across missing data mechanisms in the Absent in One Arm setting. \\

For the ML-based approaches, we observe slight bias under the Different Shapes and Absent in One Arm settings when missingness depends on baseline covariate. The magnitude of bias is reduced for MI-RF and MI-CART compared to the non-ML approaches. Across all relationships and missingness mechanisms, we observe under-estimated model-based SEs. There is slight undercoverage across the Absent in One Arm settings for MI-RF and MI-CART. We observe that MSE is reduced for MI-RF and MI-CART compared to the non-ML approaches for the Absent in One Arm setting. For MI-SL, we observe large undercoverage in all settings except when missingness depends on treatment.\\

Similar patterns are observed for smaller sample sizes (see Supplementary File 2).

\begin{figure}[]
\centering
\includegraphics[width=1\textwidth]{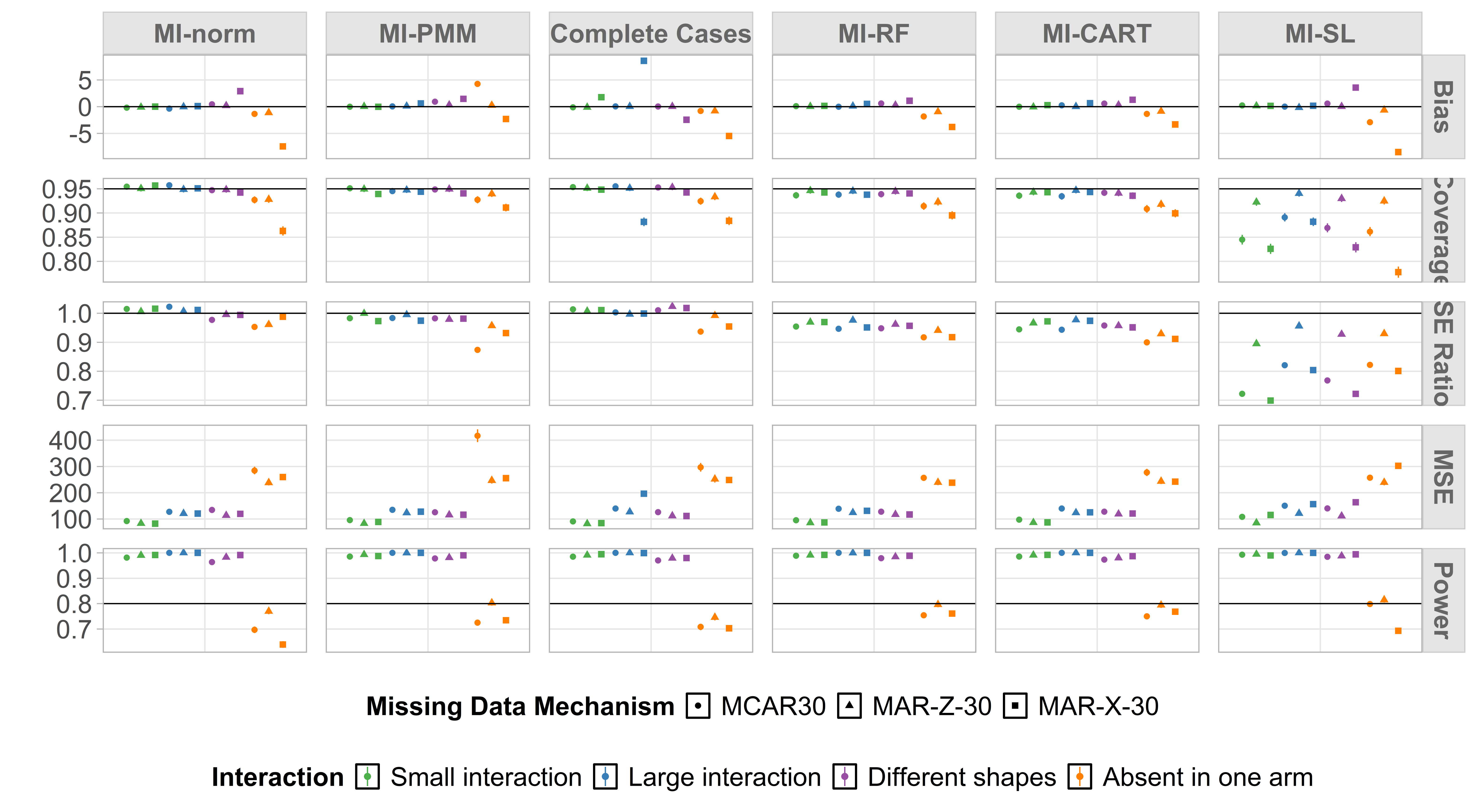}
\caption{Simulation results for Interaction setting under the alternative when sample size is 200. Performance of missing data methods in terms of bias, coverage, ratio of model-based versus empirical standard error and power are shown. Note: for RF, the number of trees is set to 5 and for CART, minbucket is set to 5. Estimates are indicated with $\pm 1.96 \times $ Monte Carlo error bars (but are often too small to be seen due to the scale of the plots). }
\label{interaction_200}
\end{figure}

\section{Simulation Study: Repeated Measures with Skewness}
\label{Section_sim_3}
The second simulation study is motivated by the 2019 iNO-PF trial by Bellerophon. This was a Phase II trial which evaluated whether inhaled nitric oxide improved physical activity in patients with Pulmonary Hypertension associated with Interstitial Lung Disease \citep{King2022}. The trial randomised 44 participants in a 2:1 ratio to treatment versus placebo. Average time spent in Moderate-to-Vigorous Physical Activity (MVPA), measured by an Actigraph wrist-worn accelerometer, was one of the exploratory outcomes \citep{King2022}. From here on, we refer to this as the \textit{MVPA} outcome. Participants wore the device for one month in an initial baseline run-in period and for a one-month follow-up four months after treatment initiation.  The Phase II trial demonstrated that MVPA increased statistically significantly in the treatment group compared to the control group; therefore, this outcome was approved as a primary outcome in the Phase III trial. However, there was no statistically significant evidence of increase in MVPA in the Phase III trial and it was therefore stopped early for futility \citep{Belle}.\\

The MVPA outcome had missing values. A valid day was defined as a day with least 600 minutes of wear time while awake, and a valid month was defined as a month with at least 14 valid days \citep{King2022}. Analysis of the Phase II trial proceeded with Complete Cases including only individuals with valid months. \\

Our simulation study is loosely based on the Phase II trial, but we simulate a larger sample size ($n=145$) as the original sample of 44 will may lead to problems with estimation (see Table 2 in Supplementary File 1), where the majority of failed simulations occur when sample size is 50.

\subsection{Aim}
The aim of this simulation study based on the iNO-PF trial is to compare performance of missing data strategies explored in the previous simulation in a more complex setting where there are (i) repeated measures and (ii) skewness in the outcomes. 

\subsection{Data Generating Mechanism}
In each repetition, we generate data for the physical activity outcome over 28 days at follow-up for 145 participants. Participants are randomised to treatment or control with a 2:1 ratio, as was done in the iNO-PF trial. We denote by $y_{ijk}$ the daily time spent in MVPA at follow-up for patient $i$ on day $j$ in week $k$, for $i \in \left\{ 1, 2, ..., 145 \right\}$, $j \in \left\{1, 2, ..., 7\right\}$ and $k \in \left\{1, 2, 3, 4 \right\}$.  \\

Time spent in MVPA has a right-skewed distribution \citep{Xue2020}; we therefore generated $y_{ijk}$ with a log-normal distribution with mean $\mu_{i}$ and standard deviation of 46, a value similar to what was observed in the trial. Note that the mean and standard deviation of the log-normal distributions are distinct from the location and scale parameters. We model $\mu_{i}$ as:

\begin{equation}
\label{mean_model}
\begin{split}
\mu_{i} &= X_i+ \delta  Z_i + u_{i}
\end{split}
\end{equation}

where 

\begin{itemize}
\itemsep0em 
\item $X_i$ is the average of the daily time spent in MVPA during the baseline period of one month, which we assume has a log normal distribution with mean 77 and standard deviation  52. These values were selected to be similar to the mean and standard deviation of the trial participants \citep{King2022}. 
\item $\delta$ is the treatment effect and is assumed to be 12, a value similar in magnitude to the effect observed in the trial \citep{King2022}. For ease of interpretation, the simulation is set up such that the experimental group experiences an increase in MVPA, while the placebo group experiences no change. This is in contrast to the iNO-PF trial, where the experimental group maintains their baseline activity level while the placebo group experiences deterioration. 
\item $u_{i} \sim N(\mu=0, \sigma^2=4)$ is a random effect for participant $i$.
\end{itemize}

This model for daily step count is simplified and focuses on its skewness characteristic; see Figure \ref{MVPA_histograms} in the Appendix for an illustration of time spent in MVPA aggregated at the day-, week- and month-level. A more realistic model may account for autocorrelation between daily measurements, zero-inflation, as well as right-truncation as daily time spent in MVPA may have an upper limit specific to the population. \\

Missing Data Mechanism: 

We consider three simple missing data mechanisms, each of which results in $30\%$ of the sample having missing outcomes: 
\begin{itemize}
    \item MCAR: $30\%$ of the sample who are selected completely at random; 
    \item MAR-X: $10\%$ of the sample with above-average baseline MVPA are selected, and $20\%$ of the sample with below-average baseline MVPA are selected. 
    \item MAR-Z: $10\%$ of the sample receiving the  experimental treatment are selected, and $20\%$ of the sample receiving placebo are selected.
\end{itemize}

For those selected participants, the first, second, third and fourth weeks have probabilities 0.25 0.5, 0.75, 0.85 of being missing, respectively. This reflects greater tendency for later weeks to be missing. We note that missing data from accelerometers can be handled at a more granular level, such as at the day or epoch-level  \citep{Di2022, tackney2021, tackney2023}. We simulate and handle missingness at the coarser week-level to focus attention on the impact of skewnesss. \\

\textbf{Estimands and analysis methods}\\

We describe below two analysis models and corresponding estimands, which were explored in \cite{King2022}. We omit additional adjustment for baseline stratification variables in these analyses for simplicity. \\

The primary analysis is an Analysis of Covariance (ANCOVA) where the outcome is the average time spent in MVPA over the month. In the iNO-PF study, a day is considered a \textit{valid} day if there is at least 600 minutes of wear time while awake; otherwise, the day is considered missing. We denote by $\bar{y}_{i..}$ the average of the outcome across valid days during the follow-up period:

\begin{equation}
\bar{y}_{i..} = \sum_{j=1}^{4} \sum_{k=1}^{7} y_{ijk} I(\mbox{$y_{ijk}$ is valid}),
\end{equation}

The primary analysis model is given by:

\begin{equation}
\label{ANOVA_analysis_model}
\bar{y}_{i..} = \beta_0 + \beta_1 Z_i + \beta_2 X_i + \epsilon_i,
\end{equation}
where we assume $\epsilon_i \sim N(0, \sigma^2)$. The estimand of interest is the average treatment effect, $\beta_1$. We set the true value to 12. \\

A second analysis is a Mixed Model for Repeated Measures (MMRM) with baseline physical activity as a covariate and treatment group, week, treatment-by-week interactions as fixed effects.  We denote by $\bar{y}_{ij.}$ the average of the valid days for participant $i$ in week $j$:

\begin{equation}
\bar{y}_{ij.} = \sum_{k=1}^{7} y_{ijk} I(\mbox{$y_{ijk}$ is valid}).
\end{equation}

The model includes baseline, week, treatment, treatment-by-week interaction as fixed effects and models correlation in the residual errors:

\begin{align}
\label{analysis_model}
\bar{y}_{ij.} &= \beta_0 + \beta_1 X_i  \nonumber \\
& + \beta_2 \mbox{I}(j=2) + \beta_3 \mbox{I}(j=3) + \beta_4 \mbox{I}(j=4) + \beta_5\mbox{treat}_i \nonumber \\
&+\beta_6 \mbox{treat}_i \mbox{I}(j=2) + \beta_7 \mbox{treat}_i \mbox{I}(j=3) + \beta_8 \mbox{treat}_i \mbox{I}(j=4) + \epsilon_{ij.}, \nonumber
\end{align}
where $\epsilon_{i} =[\epsilon_{j1.}, ..., \epsilon_{j4.}]^\top \sim N(0, \Sigma)$ are residual errors with an unstructured covariance matrix $\Sigma$ to account for the within-subject correlation.  \\

For MMRM, the estimands of interest are the treatment effects for each week (captured by the sum of the main treatment effect and interaction effect for that week) as well as the treatment effect collapsed across the four weeks (captured by the main treatment effect and average across the interaction effects). The true values of these estimands are 12. \\

\subsection{Missing data methods}
For the ANCOVA, the following methods are used to handle missing data: Complete Cases, MI-norm, MI-PMM, MI-CART and MI-RF, as described in the previous simulation. \\

For MMRM, the same MI methods are implemented. We do not use Complete Cases as this approach would discard the partially observed data from participants who provide data only on some (and not all) weeks. Instead, MMRM accounts for partially observed data by the likelihood estimation method \cite{verbeke2000linear, little2019statistical}. We refer to this approach as \textit{MMRM default}.  \\

For both the ANCOVA and MMRM analyses, missing data is handled at the week level. We perform MI separately by arm and include repeated measures as distinct variables in the imputation model (i.e. in wide format). This allows for an unstructured correlation matrix among the repeated measures \citep{Wijesuriya2024}; weeks with missing outcomes are imputed iteratively based on observed outcomes from other weeks and baseline covariate via chained equations \citep{vanBuuren2011}. We use $M=30$ for all imputation methods and use the default tuning parameter for the ML based approaches.\\

\subsection{Performance Measures}
The same performance measures are used as described in Section \ref{performance_measures}.

\subsection{Results}

\begin{figure}[]
\centering
\includegraphics[width=1\textwidth]{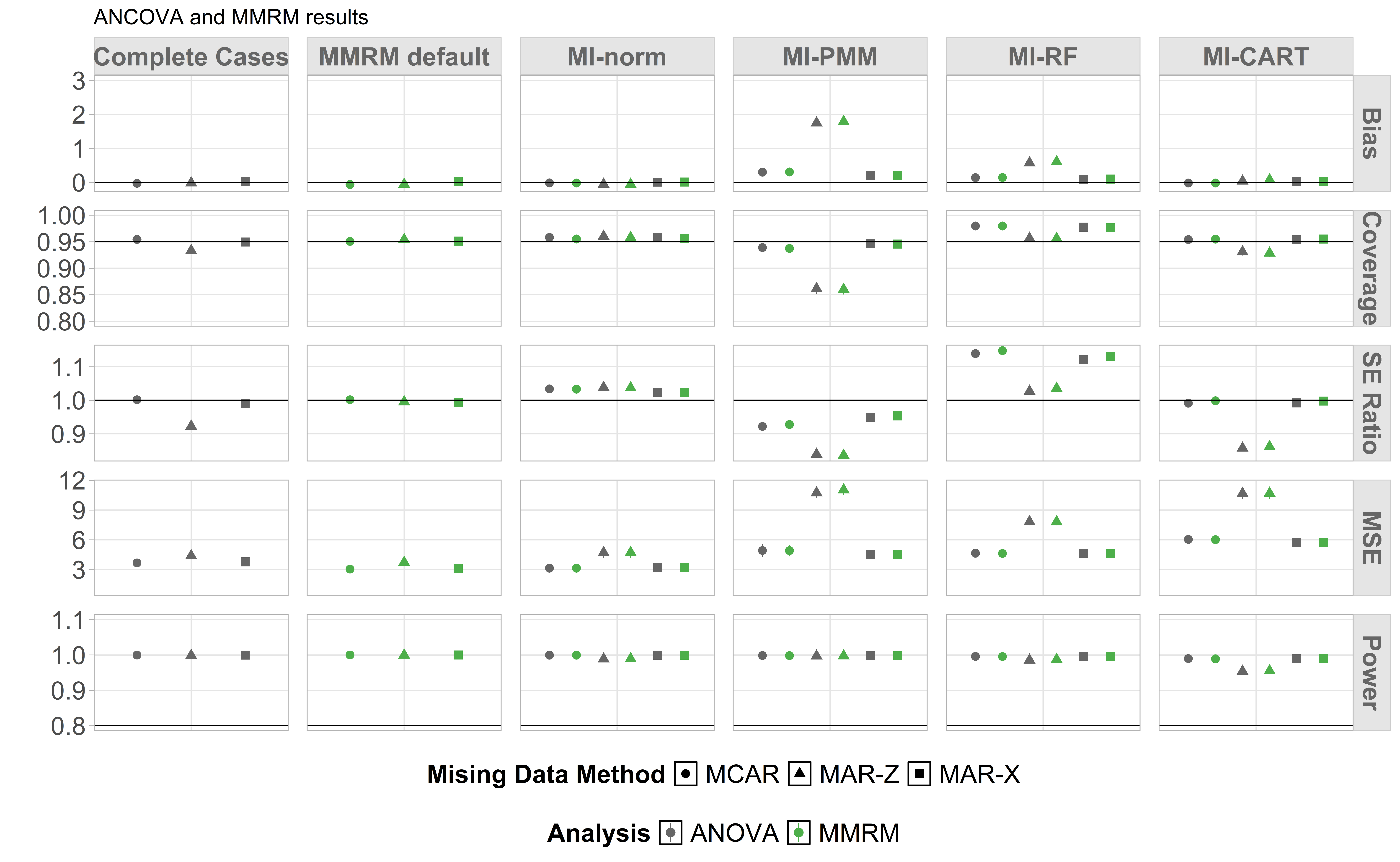}
\caption{Simulation results for the iNO-PF trial analysed with the ANCOVA and MMRM under three missing data mechanims. Performance of missing data methods in terms of bias, coverage, ratio of model-based vs empirical standard error and power are provided across seven different missing data mechanisms. Estimates are indicated with $\pm 1.96 \times $ Monte Carlo error bars (but are often too small to be seen due to the scale of the plots). }
\label{ANCOVA_MMRM}
\end{figure}

Figure \ref{ANCOVA_MMRM} displays results for the ANCOVA and MMRM. Comparing the results of ANCOVA with Complete Cases with MMRM default, we observe that estimates are unbiased. There is slight undercoverage and under-estimation of SEs for ANCOVA with Complete Cases under MAR-Z, while MMRM default has nominal coverage and SE ratio at 1. Furthermore, MSE is slightly reduced for MMRM default compared to ANCOVA Complete Cases for all missing data mechanisms. This is likely due to the fact that MMRM default uses available data from all individuals whereas ANCOVA with Complete Cases discards data from individuals whenever any weekly outcome is missing. \\

Under the MI approaches, ANCOVA and MMRM provide identical results. This is expected, as the ANCOVA and MMRM which includes interactions between the baseline covariate and time point give identical point estimates when data are complete or are imputed \citep{Schuler2021}. \\

For MI-norm, we observe unbiased estimates and nominal coverage; SEs are slightly over-estimated and we observe that MSE are higher for MAR-Z compared to the MCAR and MAR-X. MI-PMM leads to bias, undercoverage and underestimated SEs across missing data mechanisms, particularly for MAR-Z.\\

For the ML approaches, we observe that MI-RF leads to bias under MAR-Z and inflated SEs and overcoverage under MCAR and MAR-X. MI-CART is unbiased across the scenarios but led to underestimated SEs and hence undercoverage under MAR-Z. For MSE, ML approaches and MI-PMM have comparable values to each other and are larger than the MSE for ANOVA Complete Cases, MMRM default and MI-norm. \\  

 Figures \ref{ANCOVA_means} and \ref{MMRM_means} in the Appendix display estimated means by treatment arm for the ANCOVA and MMRM, respectively. We observe that estimated means are are all biased upwards, probably due to the skewness in the outcome data that cannot be captured by the methods that assume normality. Nevertheless, in many cases, the biases are in equal magnitude for the two arms, which then cancel out when computing the treatment effect estimates. Exceptions are for MI-PMM and MI-RF under MAR-Z. In particular for MI-PMM,  bias is smaller for the control arm compared to the treatment arm for ANCOVA, and bias is reduced for weeks 2-4 for the control arm for MMRM. For MI-RF, we observe smaller bias for estimated means for weeks 3-4 for the control arm for MMRM. These led to bias in the estimated treatment effect for MI-PMM and MI-RF. \\

\section{Discussion}
\label{Section_discussion}

This investigation is, to our knowledge, the first comparison of parametric versus machine-learning approaches to MI in the RCT setting. While comparisons have previously been conducted in observational settings, there are distinct features of RCTs that warrant this investigation. Firstly, RCTs generally have smaller sample sizes than observational settings, which may limit potential advantages of ML approaches. Secondly, RCTs have simpler analysis models and typically adjust for the randomised arm, covariates used in stratified randomisation and pre-selected covariates to improve precision. Therefore, since there is potential for misspecification of both the imputation and analysis models, the flexibility of ML approaches for imputation may be attractive. Third, for Phase III trials, strict Type I error control is usually needed to meet regulatory requirements; therefore, the performance of MI approaches under the Null hypothesis require investigation as comparisons in the observational settings have generally focused performance under the Alternative hypothesis. \\

We summarise the conclusions from the two simulation studies which demonstrated performance of parametric and ML approaches to MI, as well as Complete Cases, in a number of settings where there is misspecification. We also outline areas of future work. 

\subsection{Conclusions from Simulation Studies}

We found that Complete Cases led to estimators with good properties in the absence of treatment-covariate interactions. In the presence of interactions, Complete Cases led to bias in estimating the ATE when missingness depended on baseline covariate; proof of this result is provided in Supplementary File 3. \\ 

In the absence of interactions, MI-norm had good performance except when missingness depended on the baseline covariate. While \cite{Sullivan2018} demonstrated good performance of MI-norm when the analysis model is correctly specified, we demonstrate that MI-norm can lead to reduced coverage and under-estimated standard errors under model misspecification. MI-PMM led to bias, poor coverage, inflated type I errors and under- or over-estimated SEs in a large number settings. Our simulations used the default value of 5 donors for MI-PMM; when there is potential for non-linear relationships, the size of donor pool may need to be larger. In the presence of simpler treatment-covariate interactions, MI-norm and MI-PMM performed well, but displayed bias and undercoverage for more complex interactions when missingness depended on the baseline covariate. Our investigation builds on findings by \citep{Sullivan2018}, who showed that imputing separately by arm produced unbiased treatment effects when the outcome model omits an interaction. We further show that, when there are non-linearities in the covariate-outcome relationship in addition to an omitted interaction, imputing separately by arm can lead to bias.\\

MI-RF and MI-CART generally had low bias in the absence of treatment-covariate interactions, except when missingness depended on treatment. For simpler covariate-outcome relationships, these approaches lead to inflated type I error, undercoverage and underestimated SEs. However, in highly non-linear settings at sample sizes of 200 or larger, we found that they led to reduced MSE and gain in power compared to non-ML approaches while maintaining good coverage and type I error. When there is a simpler covariate-outcome interaction, we observed that MI-RF and MI-CART had similar performance to MI-norm and MI-PMM. For more complex interactions, MI-RF and MI-CART had improved coverage and reduced MSE compared to non-ML approaches. MI-SL displayed some bias, inflated type I error and large under-coverage, across a number of covariate-outcome relationships and interaction settings. \\

Overall, no single approach had superior performance across all settings in the single outcome simulation. The choice of missing data approach may depend on the potential for complexities in the underlying relationship, the sample size of the trial as well as whether purpose of the trial is for decision making or more exploratory. For late-phase trials where strict control of type I error rate is needed, Complete Cases may be more appropriate than MI under the conditions explored in our simulations, as they lead to Type I error controlled at the nominal level. If MI is preferred (for example to include auxiliary variables), MI-norm was better able to keep Type I error controlled at the nominal level across a greater number of settings. For earlier-phase trials that are geared toward hypothesis generation, MSE may be an important metric and the potential for ML approaches to lead to decreased MSE in the presence of non-linearities and complex interactions may be attractive. However,  an outstanding issue with ML approaches to MI is that Rubin’s Rules for pooling standard errors generally led to underestimated standard errors. \\

We investigated the impact of skewness of the outcome in a trial setting motivated by digitally measured physical activity outcomes. Here, the outcome can either be summarised at the weekly or monthly level, and analysed using an MMRM or ANCOVA, respectively. We observed that the estimated means for were generally biased, but the treatment effect was unbiased except when missingness depended on treatment for MI-PMM and MI-RF. MMRM default appeared to have ideal performance across missingness scenarios considered, despite skewness in the outcome, whereas MI led to under- or over-coverage, reduced or increased SEs and increased MSE. \\

\subsection{Future work}

There are several areas of future work to enable comprehensive guidance on when ML approaches to MI may be preferred in trial settings. Firstly, the standard justification for Rubin’s Rules is within a parametric Bayesian framework. As such, there appear to be no guarantees that Rubin's Rules are valid when machine learning methods are applied in MI. Moreover, related results from so-called debiased machine learning suggest that existing imputation estimators based on machine learning will not have the usual asymptotic properties that parametric methods possess (under correct model specification) \citep{Chernozhukov2018}. Development of ML approaches to MI with alternative standard error estimators that provide correct inference is a key area of future work. This will have important implications as standard errors are often used in meta-analyses as well as in planning of future trials. \\

Secondly, our results showed that, in the presence of non-linear interactions, all approaches lead to some bias under MAR-X. Literature on covariate-adjustment approaches that allow for interactions, such as G-computation with interaction and Inverse Probability of Treatment Weighting, have shown improved performance compared to simple adjustment of covariates via an ANCOVA \citep{Tsiatis2008,Tackney2023Covariate}. Moreover, recent work has introduced the topic of using machine learning to assist with covariate adjustment, although typically under an MCAR assumption for any missing outcomes \citep{Williams2022}. Investigation of missing data methods in conjunction with more complex covariate adjustment approaches is an area of future work. \\

Thirdly, in order to provide further recommendations on the use of ML approaches to MI, extended simulation studies exploring additional settings are needed, once a reliable estimator of standard error of treatment effect is established. These include investigation of the number and nature of learners included in MI-SL. In the repeated measures setting, exploration of more complex missing data mechanisms are warranted, where missingness in later weeks may depend on values of outcomes in earlier weeks. In the specific setting of accelerometer outcomes, using ML approaches to handle missing data at the finer, epoch level \citep{Di2022, tackney2023}, rather than on the summary level (as was explored in our simulation), is an area which needs further investigation. \\

Fourth, our investigation considered settings where outcomes are missing but covariates are fully observed. Further investigation is needed to in settings where both outcomes and covariates are missing. Particularly when considering treatment-covariate interactions with missing values in covariates, the implications of using ML versus parametric approaches to MI for subgroup analyses in precision medicine, for example via innovative designs such as Sequential Multiple Assignment Randomized Trials \citep{Shortreed2014}, may be an interesting avenue of future work. \\
\newpage

\section{Appendix}

\begin{figure}[]
\centering
\includegraphics[width=0.7\textwidth]{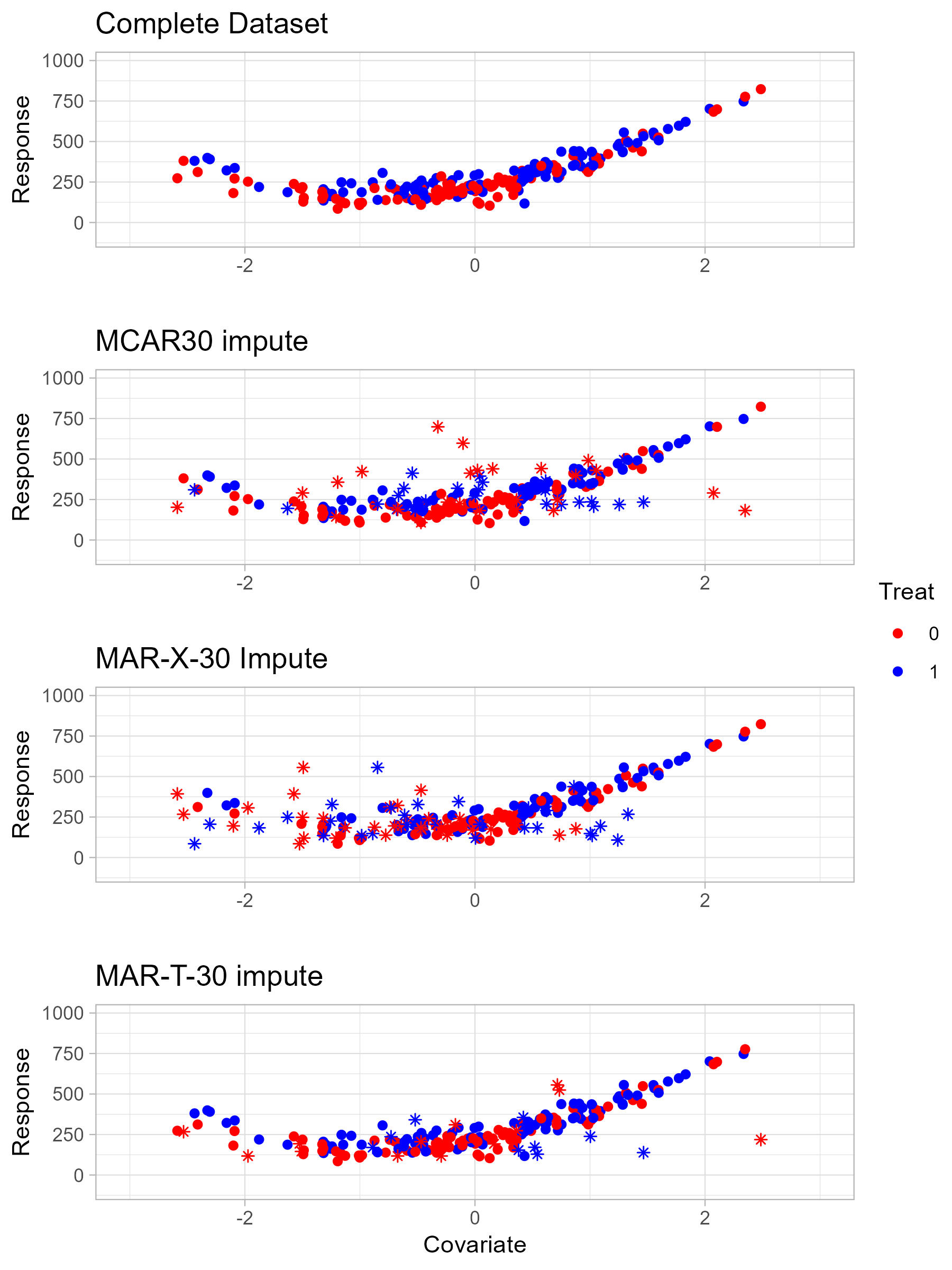}
\caption{Visualisation of imputations for the J-shaped relationship for when the true treatment effect is 40 and sample size is 200. We illustrate the fully observed dataset on the top panel. The other panels illustrate the dataset where outcomes are made missing under the specified missing data mechanism and then imputed (shown by a star) using PMM-default. }
\label{jshaped_imputations}
\end{figure}

\begin{figure}[]
\centering
\includegraphics[width=0.7\textwidth]{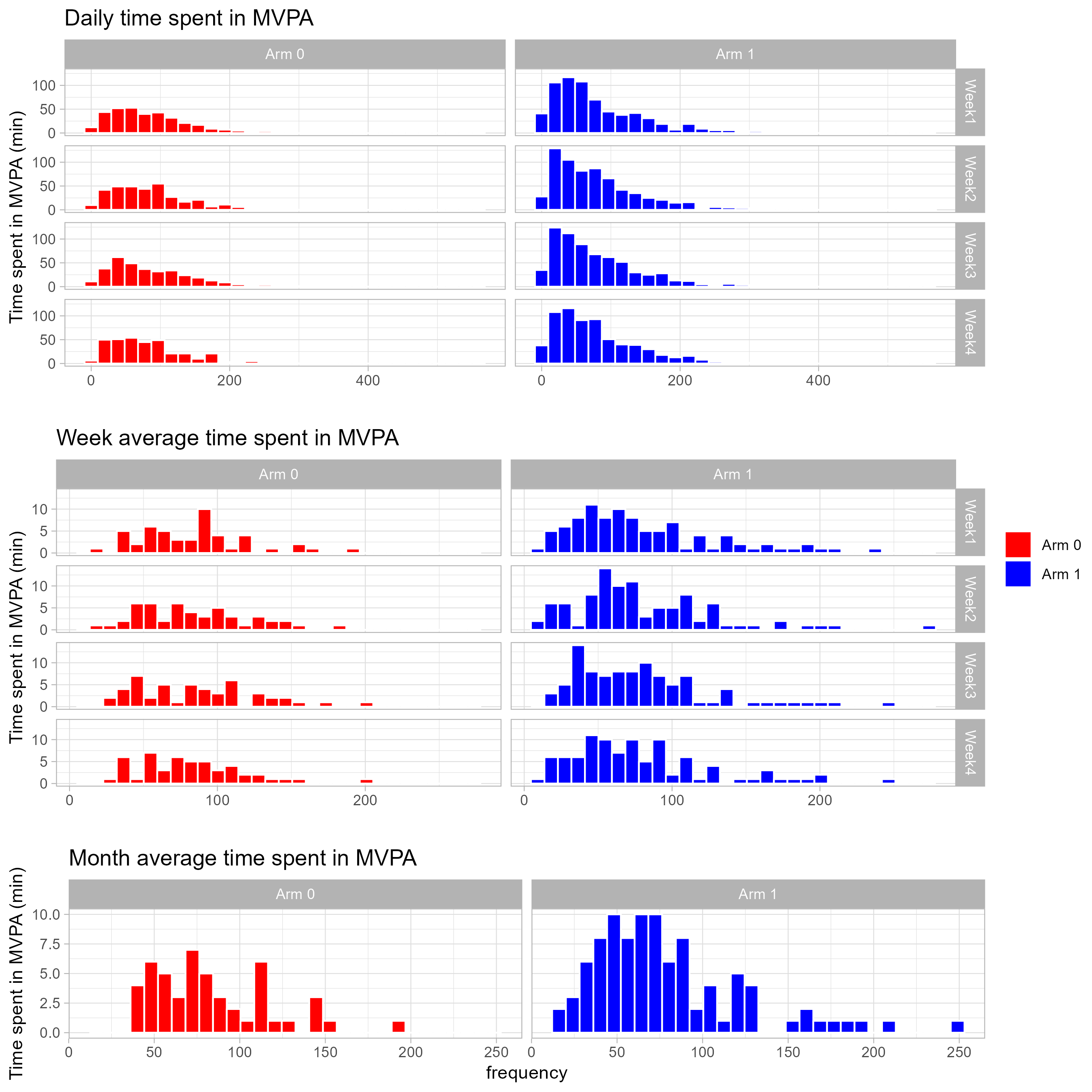}
\caption{Histograms of daily time spent in MVPA (top panel), the average daily time spent in MVPA per week (middle panel) and the average daily time spent over the entire month-long measurement period (bottom panel) for a simulated dataset for the iNO-PF trial simulation study. As data are aggregated, we observe that outcomes become less skewed. }
\label{MVPA_histograms}
\end{figure}

\begin{figure}[]
\centering
\includegraphics[width=1\textwidth]{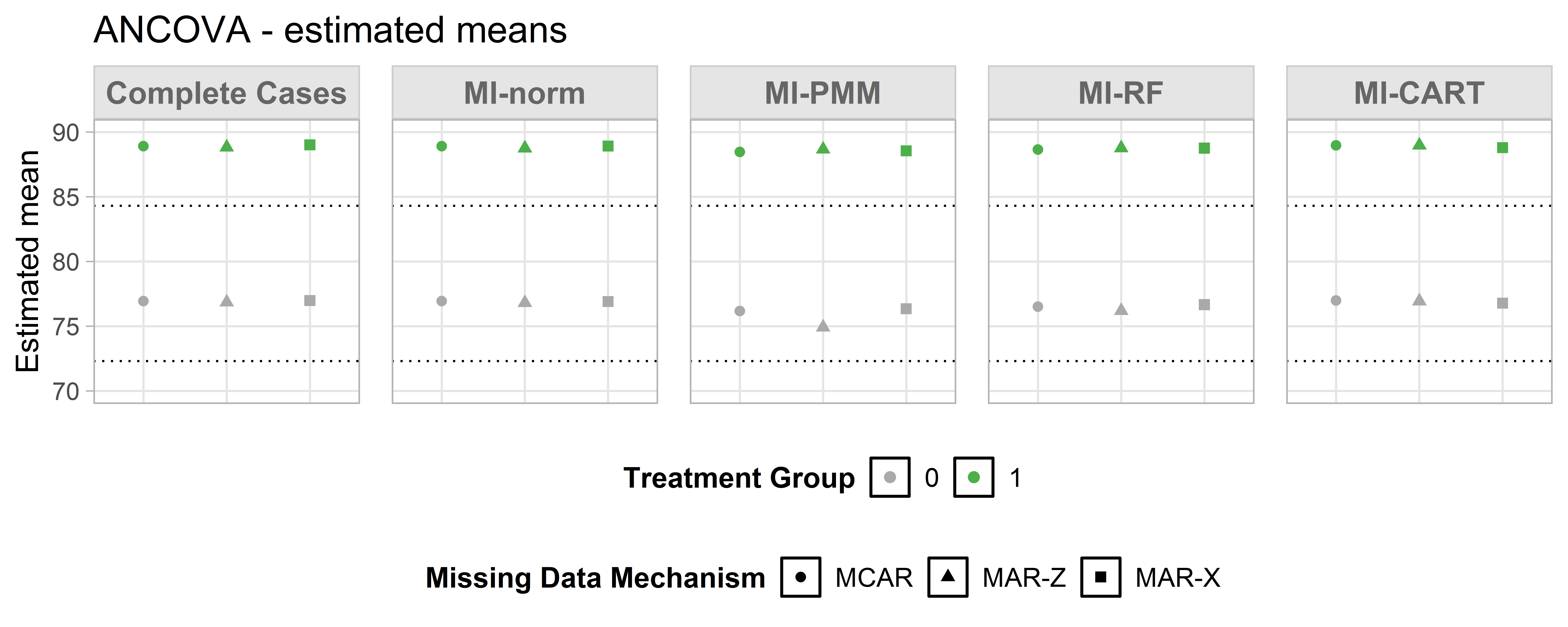}
\caption{Simulation results for the iNO-PF trial analysed with ANCOVA. Results for the estimated mean outcome for each treatment is shown under different missing data mechanisms and missing data methods. The dotted line indicates the true outcome for each treatment.  }
\label{ANCOVA_means}
\end{figure}

\begin{figure}[]
\centering
\includegraphics[width=1\textwidth]{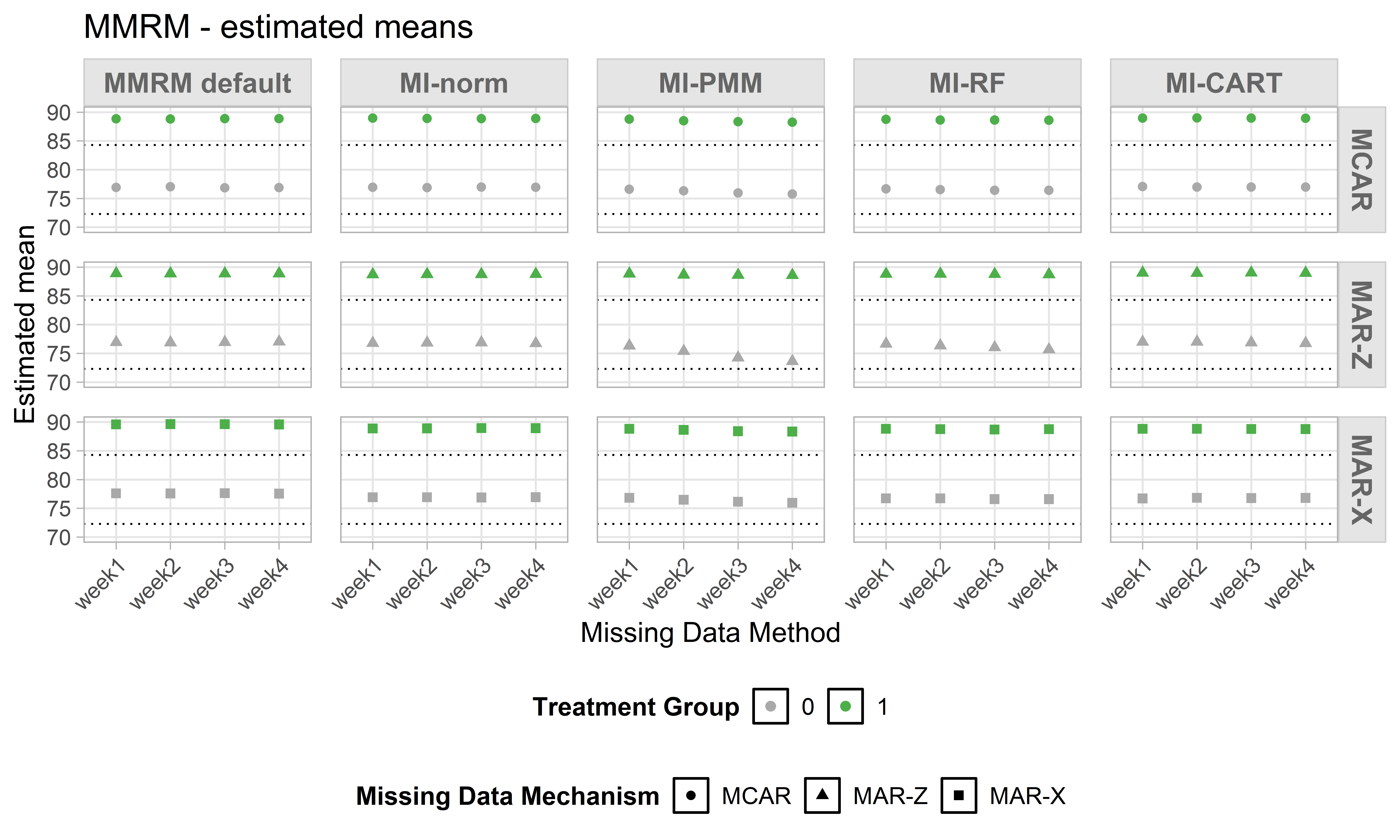}
\caption{Simulation results for the iNO-PF trial analysed with MMRM. Results for the estimated mean outcome for each treatment for each week is shown under different missing data mechanisms and missing data methods. The dotted line indicates the true outcome for each treatment. }
\label{MMRM_means}
\end{figure}

\section*{Declaration of Conflicts of Interest}
JWB and his employers have received fees for statistical consultancy
from AstraZeneca, Bayer, Novartis, and Roche.

\section*{Funding}
MST, Advanced Fellow, NIHR305417, is funded by the National Institute of Health and Care Research for this research project. KML is supported by the National Institute for Health and Care Research (NIHR300051). The views expressed are those of the authors and not necessarily those of the NIHR or the Department of Health and Social Care. EW was supported by a Wellcome Senior Research Fellowship (224485/Z/21/Z).

\section*{Supplementary Materials}
All supplementary files are provided in \url{https://github.com/mst1g15/MICE-comparison}. Supplementary File 1 contains additional details and results of the simulation study. Supplementary File 2 contains simulation results for sample sizes not presented in the main paper. Supplementary File 3 contains a proof demonstrating that Complete Cases yields valid inference under conditions explored in our simulations. Code to run the simulation, and plots of all results, are available on the GitHub repository.\\

\newpage
\bibliographystyle{plainnat}

\bibliography{refs}

\end{document}